\documentclass[apjl]{emulateapj}
\usepackage[]{times, graphicx}
\newcommand{\hinode}{{\em Hinode}}
\newcommand{\trace}{{\em TRACE}}

\newcommand{\pref}{\protect\ref}

\begin{document}

\shorttitle{Dynamic Coronal Roots}
\shortauthors{McIntosh \& De~Pontieu}
\title{Observing Episodic Coronal Heating Events Rooted in Chromospheric Activity}

\author{Scott W. McIntosh\altaffilmark{1}, Bart De Pontieu\altaffilmark{2}} 
\altaffiltext{1}{High Altitude Observatory, National Center for Atmospheric Research, P.O. Box 3000, Boulder, CO 80307}
\altaffiltext{2}{Lockheed Martin Solar and Astrophysics Lab, 3251 Hanover St., Org. ADBS, Bldg. 252, Palo Alto, CA  94304}
\email{mscott@ucar.edu, bdp@lmsal.com}

\begin{abstract}
We present results of a multi-wavelength study of episodic plasma injection into the corona of AR 10942. We exploit long-exposure images of the \hinode{} and Transition Region and Coronal Explorer (TRACE) spacecraft to study the properties of faint, episodic, ``blobs'' of plasma that are propelled upward along coronal loops that are rooted in the AR plage. We find that the source location and characteristic velocities of these episodic upflow events match those expected from recent spectroscopic observations of faint coronal upflows that are associated with upper chromospheric activity, in the form of highly dynamic spicules. The analysis presented ties together observations from coronal and chromospheric spectrographs and imagers, providing more evidence of the connection of discrete coronal mass heating and injection events with their source, dynamic spicules, in the chromosphere.
\end{abstract}


\section{Introduction}
Observing the mechanics of the process that raises the 10,000 K chromospheric plasma to several million K has eluded the community since the inference of hot coronal plasma was made at the dawn of the rocket age \citep{Edlen1943}. Recently, analysis of data from the \hinode{} spacecraft \citep[][]{Hinode} has established the connection of chromospheric and coronal heating processes \citep{DePontieu2009}. The key to this observational-driven leap in understanding is the discovery of a second class of ``spicule'' \citep[e.g.,][]{Roberts1945,Beckers1968}. These so-called type-II spicules were revealed in the high spatial resolution observations of the chromospheric limb provided by the Solar Optical Telescope \citep[][]{Tsuneta2008}. \citet{DePontieu2007b} showed that they originate in strong magnetic regions, are longer (4-8Mm), and display larger upward velocities (50-150km/s) than their classical (shock-driven) ``Type-I'' counterparts \citep[2-5Mm tall with speeds 10-40km/s;][]{DePontieu2004,Hansteen2006,DePontieu2007a,Rouppe2007}. 

\citet{DePontieu2009} demonstrated the spatio-temporal correlation of upper chromospheric activity, in the form of Type-II spicules, and a weak component (2-5\% of the peak line emission) in the blue wings of three emission lines formed in the transition region and corona (\ion{C}{4} 1548\AA{} at 150,000K, \ion{Ne}{8} 770\AA{} at 600,000K, and \ion{Fe}{14} 274\AA{} at 2MK). This correlation was strongest in locations of unipolar plage, as well as the supergranular network vertices of the quiet sun and coronal holes. In each case the blue-wing component, indicative of the presence of an upward moving jet, shows a nearly identical velocity distribution to that of the dynamic spicules seen in the same unipolar regions. This led \citet{DePontieu2009} \citep[and][]{McIntosh2009a} to suggest that these episodic heating events rooted in chromospheric activity play a significant role in filling the Sun's upper atmosphere with hot plasma.

\begin{figure}
\epsscale{1.15}
\plotone{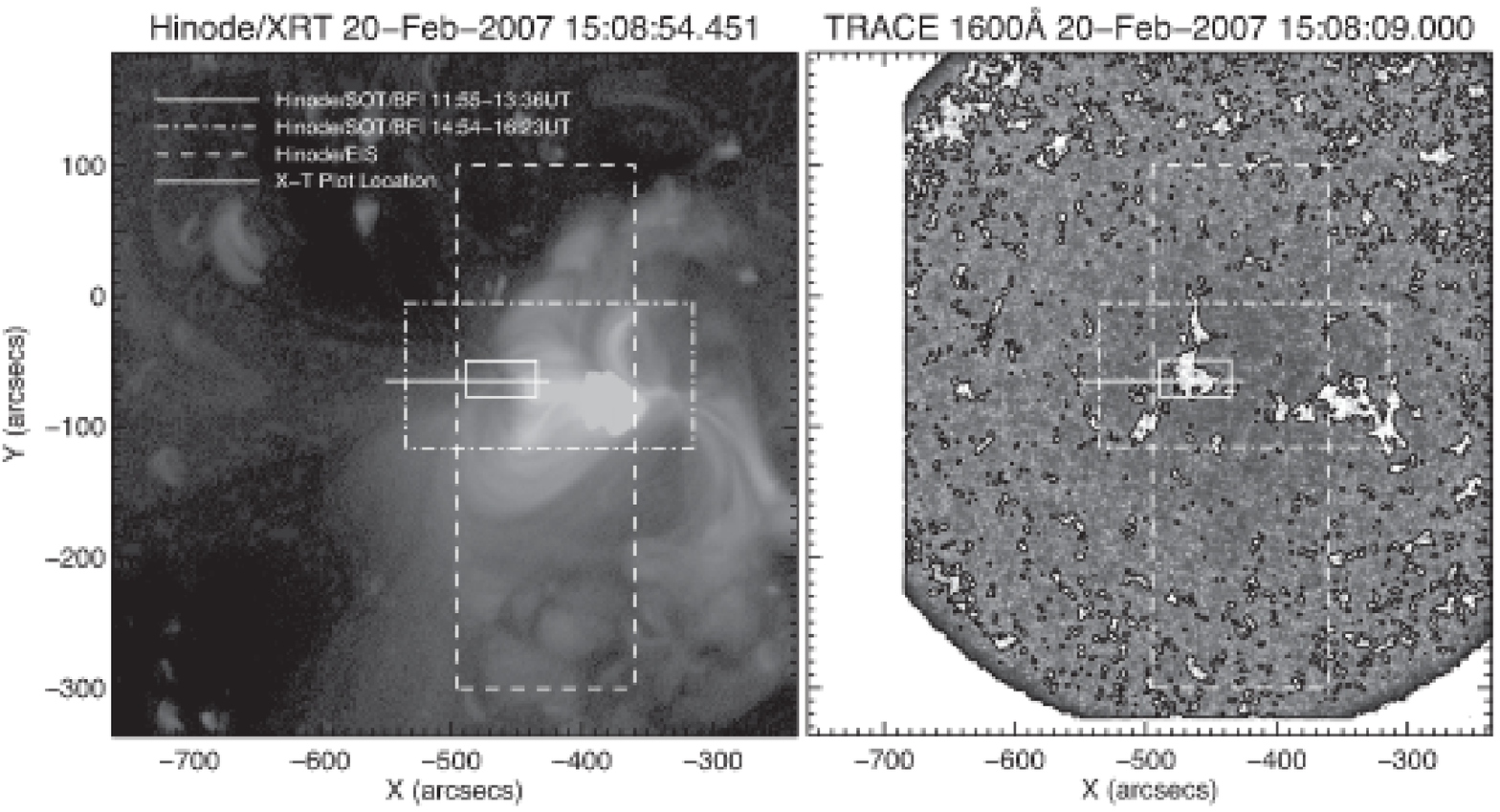}
\plotone{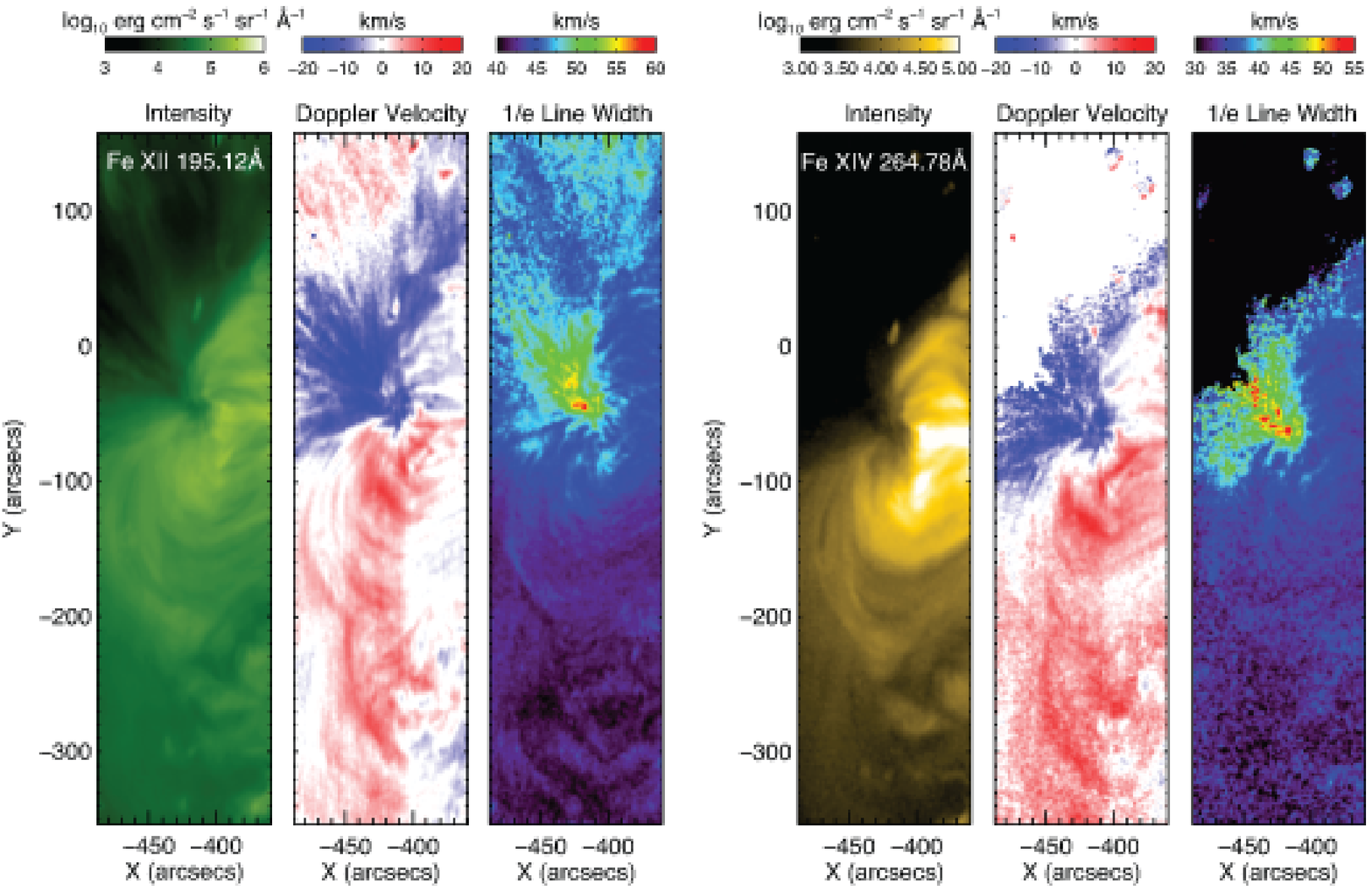}
\caption{Context imaging and spectroscopy of AR 10942 on 2007, February 20 from XRT (top left), \trace{} 1600\AA{} (top right) and EIS in the \ion{Fe}{12} 195\AA{} (bottom left) and \ion{Fe}{14} 264\AA{} (bottom right) lines. We show the EIS and different SOT fields of view for a short-exposure co-temporal sequence (dot\--dashed line) and for a longer exposure sequence taken beforehand (dot\--dot\--dashed line), and the reference position used to compute the space-time (x-t) plots used throughout this Letter. The bright plage and network emission in the TRACE image are outlined by an intensity contour of 250~DN. \label{f1}}
\end{figure}

In this Letter we explore a long-exposure observing sequence performed by \hinode{} and the Transition Region and Coronal Explorer \citep[\trace;][]{Handy1999} of solar active region (AR) 10942. Observations with the \hinode{} X-Ray Telescope \citep[XRT;][]{Golub2007} revealed the presence of episodic, fast, ``blobs'' of coronal material ($>1$MK) that appear to emanate from the plage in the active region \citep[][]{Sakao2008}. Our motivation comes from the fact that the reported blob velocities (qualitatively) matched those of Type-II spicules and associated weak coronal upflows (measured spectroscopically) observed in the magnetic footpoints of AR coronal loops. In the sections that follow we explore the thermal connectivity and origins of these coronal blobs using an extended observational dataset, taken the day before the observations discussed by \citet[][]{Sakao2008}.

\section{Observations}
The observations were taken on 2007 February 20 with the \hinode{} XRT, SOT and Extreme-ultraviolet Imaging Spectrometer \citep[EIS;][]{Culhane2007} instruments with support from \trace{} \citep[][]{Handy1999}. The observations were part of the joint observing campaign that focused on AR 10942 as it transited the solar disk \citep[for details, see][]{Sakao2008}. The fields of view of the \hinode{} instruments used here are shown in the left panel of Fig.~\pref{f1}. The location of the AR plage and network is shown with intensity contours on top of the TRACE 1600\AA{} image in the right panel. 

XRT observed the AR continuously from 11:52-17:37 in the Ti-Poly filter with a mean cadence of 116s and a range of exposures (from 1 to 16s). \trace{} observed the AR from 15:09 to 17:59 UT in its 171\AA{} passband with a mean exposure and cadence of 77s and 85s, respectively. EIS rastered a 128\arcsec{} $\times$ 512\arcsec{} region from 15:10-17:22UT acquiring data in 17 spectral windows that covered a broad range of (equilibrium) formation temperatures and a 60s exposure at each step. We select the longest exposure frames ($>$10s) in the XRT and \trace{} sequences that span the time period from 15:08-17:36UT. There are 89 and 85 frames respectively in each sequence. 

SOT observed this region twice in the \ion{Ca}{2}H filter of BFI with relatively high cadence timeseries, once with 0.128s (14:54-16:23UT; 16s cadence) and another with 0.410s exposures (11:55-13:36; 6.4s cadence). While the latter sequence is not co-temporal with the observations from the other instruments, it has exposures deep enough to allow us to isolate the faint, highly dynamic upper chromospheric emission in the BFI \ion{Ca}{2}H passband \citep[see, e.g.,][]{DePontieu2009} from the dominant photospheric contribution on the disk.

In each case we coalign the image sequences using a Fourier cross-correlation technique, performing intra-instrument coalignment in the same fashion. The SOT images of the chromosphere can be coaligned to the XRT images by using \trace{} 1600 and 171 \AA\ images with an accuracy of about 1\arcsec{} although precise coalignment is not necessary \-- our intent is to determine and compare the velocities of the blobs (and related signatures) across chromospheric and coronal temperatures in a region, co-spatial to the pointing tolerance of EIS.

\begin{figure}
\epsscale{1.15}
\plotone{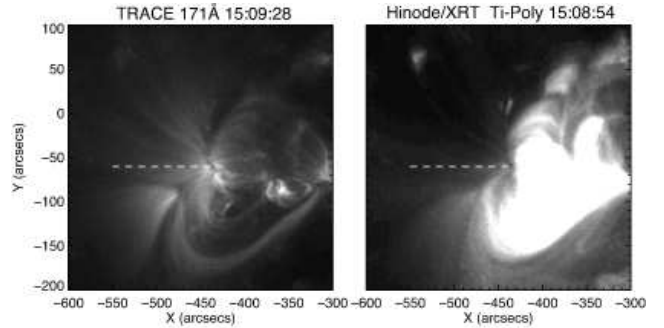}
\caption{Coaligned and (nearly) co-temporal observations from \trace{}, in the 171\AA{} passband and \hinode{}/XRT in the Ti-Poly filter configuration. The latter is saturated to enhance the contrast in the weaker eastern side of the active region. The horizontal dashed line is the reference position used to compute the space-time (x-t) plots used throughout this Letter. \label{f2}}
\end{figure}

\begin{figure*}
\epsscale{1.15}
\plotone{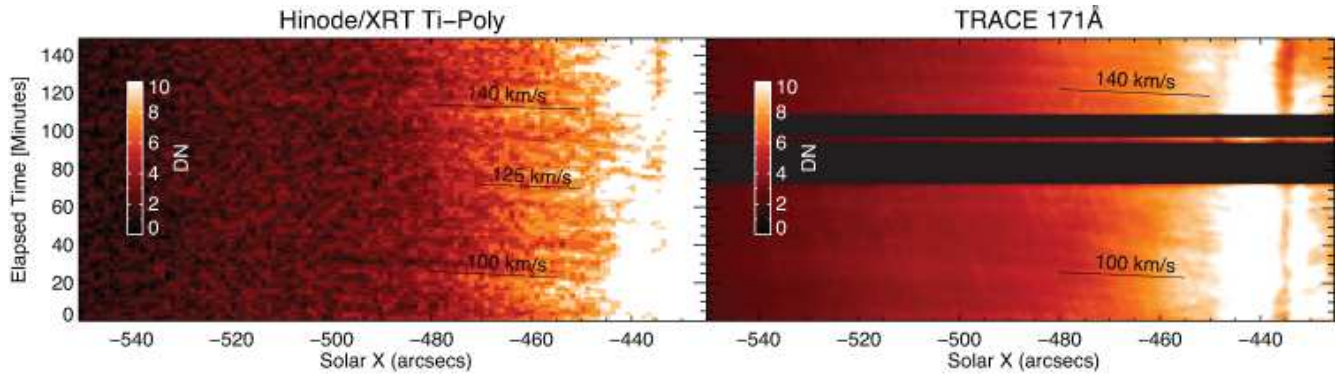}
\caption{Space-Time (x-t) plots determined from the long-exposure \hinode{}/XRT (left) and \trace{} (right) image sequences shown in Fig.~\pref{f2} interpolated to the same temporal scale. The nearly horizontal striations in the plots, running left to right are the signature of the very weak intensity blobs observed. The horizontal black bands in the \trace{} image are times when the spacecraft flies through the South-Atlantic Anomaly. \label{f3}}
\end{figure*}

\section{Analysis}

In order to analyze the blobs visible in the \trace{} and XRT sequences (Fig.~\pref{f2} and associated movie) we study space-time (x-t) plots at selected locations around their origin in the plage in a fashion similar to \citet{Sakao2008}. The dashed line shown in the two panels of Fig.~\pref{f2} (dashed line) is a sample. The data underlying the sample ``slit'' is extracted from the image sequence and interpolated onto the same timescale (60s) for ease of representation using a cubic interpolation scheme. We see the strong visual correspondence between the two x-t plots shown in Fig.~\pref{f3}, noting that the signals of the blobs in each sequence are weak (1-10DN), at most a few percent of the coronal brightness in the remainder of the AR. The \trace{} 171\AA{} blobs show a little more contrast in the region extended eastward from the plage region. Shown in each panel are a small sample of the blob velocities (straight lines in the x-t plot) showing typical values from 100 to 140km/s. The measured velocities are lower bounds - the blobs on the periphery of the plage are likely to be (highly) inclined with respect to the horizontal and so we would expect to infer a somewhat reduced speed. These x-t plots, derived velocities and blob intensities are typical of the corona above the entire plage region. 

\begin{figure}
\epsscale{1.15}
\plotone{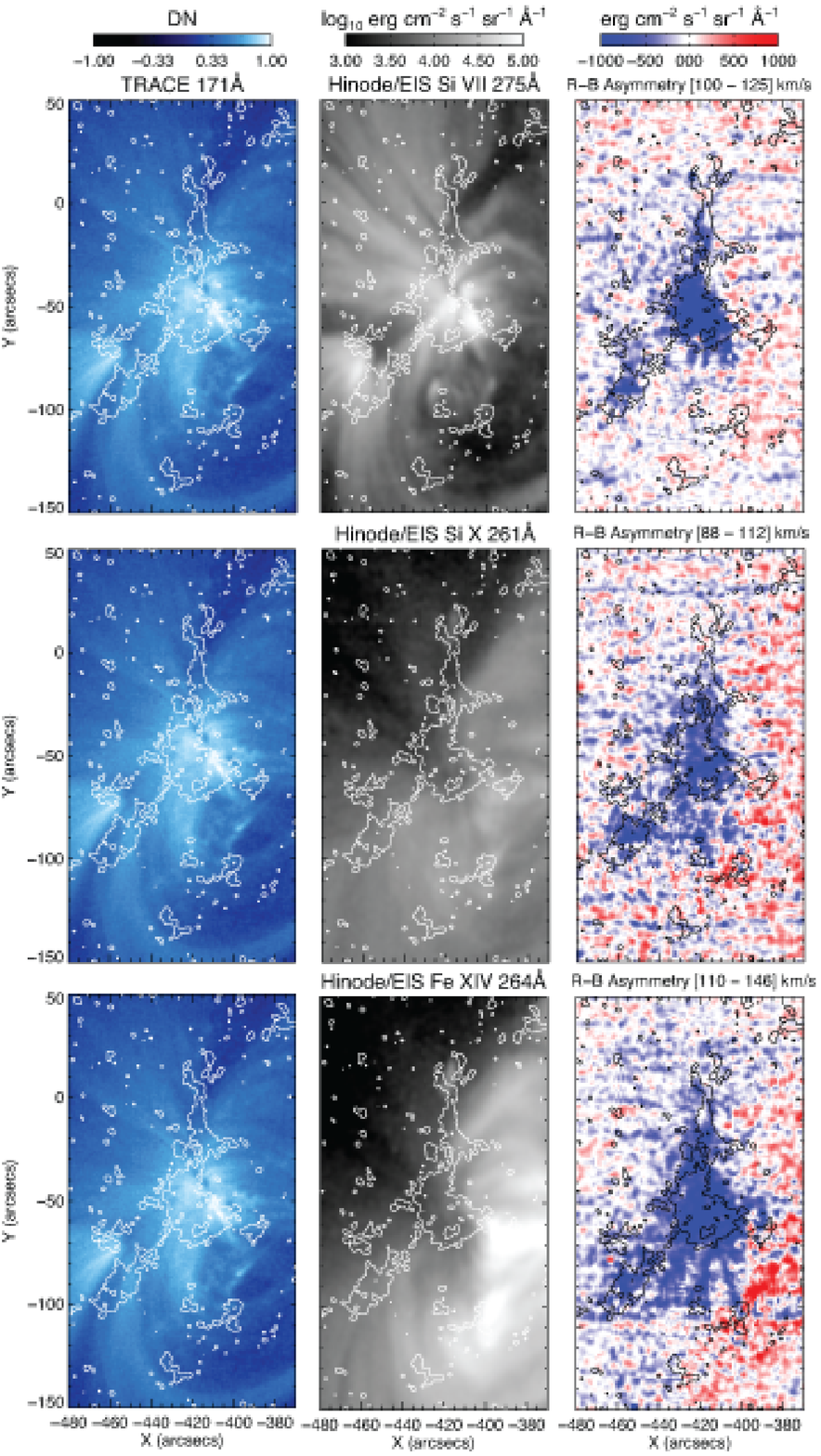}
\caption{Co-temporal \hinode{}/EIS spectro-heliograms of AR 10942 in three spectral lines: \ion{Si}{7} 275\AA{} (top row); \ion{Si}{10} 261\AA{} (middle row); \ion{Fe}{14} 264\AA{} (bottom row). Each row of the figure is comprised, from left to right, of the \trace{} 171\AA{} image at the start of the sequence (see, e.g., Fig.~\pref{f2} and associated movies), the peak intensity image, and the Red-Blue line profile asymmetry at a range of velocities (see text for details). Each panel shows the contour of \trace{} 1600\AA{} intensity to outline the plage locations (see, e.g., Fig.~\pref{f1}). \label{f4}}
\end{figure}

Figure~\pref{f4} shows the results of performing a profile asymmetry analysis on a selection of the available EIS spectral window rasters \citep[][]{DePontieu2009}. This ``R-B'' analysis involves several steps. First we fit a single Gaussian shape to the emission line profile at each pixel to establish the line centroid. Once determined, we sum the amount of emission in narrow ($\sim$24km/s wide) spectral windows symmetrically placed about that centroid in a line profile interpolated to ten times the spectral resolution. We then subtract the red and blue wing contributions to the interpolated profile (hence R-B) to make a filtergram sampling a particular velocity range, dividing the result by the factor of ten. A positive value of R-B indicates an asymmetry in the red wing of the line, which we can interpret as the signature of excess downflowing material at that velocity while, conversely, a negative value of R-B would indicate an excess of upflowing material. The rows of the figure show a region of the EIS field-of-view in the \ion{Si}{7} 275\AA{} (top row), \ion{Si}{10} 261\AA{} (middle row), and \ion{Fe}{14} 264\AA{} (bottom row) spectral windows which contain lines formed (in equilibrium) over a relatively broad range at 0.6, 1.3 and 2.0MK, respectively \citep[][]{Mazzotta1998}. In the right column of the figure we show the R-B profile asymmetry in those spectral windows at a range of velocities\footnote{The movies in the online edition of the journal show the complete set of R-B measures in each line.}. We see, using the \trace{} 1600\AA{} plage outline contours shown, that the faint fast upflows occur in the plage region at the footpoints of the coronal loops -- where the \trace{} and XRT images revealed fast propagating blobs of plasma. In addition, the broad range of velocities observed in the faint upflow emission component (determined from the R-B measure) match that inferred from the imaging data (Fig.~\pref{f3}).

\begin{figure*}
\epsscale{1.15}
\plotone{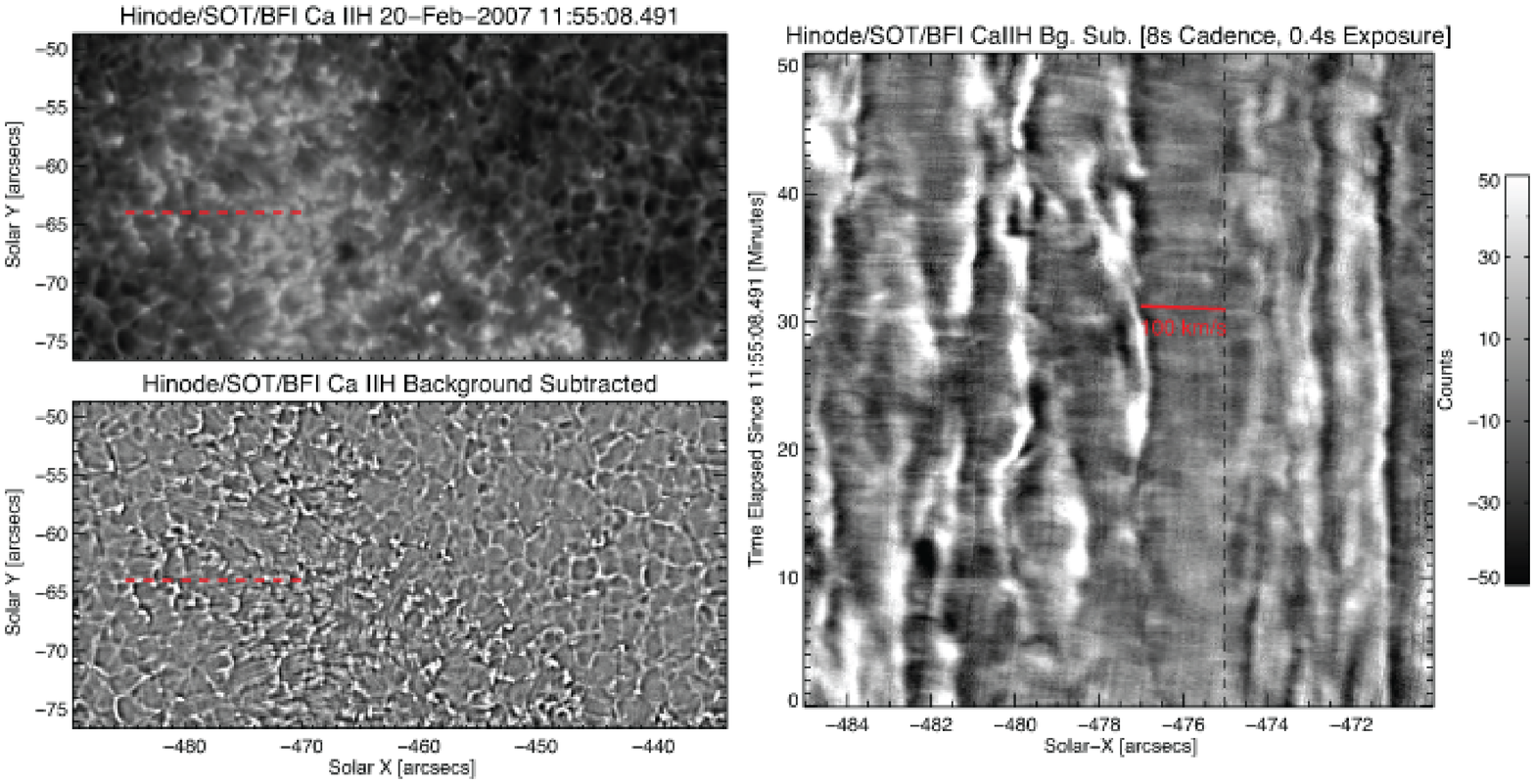}
\caption{The long-exposure \hinode{}/SOT \ion{Ca}{2}H image sequence of the following polarity plage of AR 10942. The left column of the plot shows the calibrated \ion{Ca}{2} images (top) and the background-subtracted images (bottom) derived to increase the contrast of the type-II spicules seen around the plage of the AR. Both of these figures show the reference position (red-dashed) line used above, corrected for solar differential rotation. The right column of the figure shows the background-subtracted x-t plot at that location. Shown, for reference, is a red solid line indicative of a spicule velocity of 100km/s along the line. \label{f5}}
\end{figure*}

Finally, we consider the long-exposure SOT observations of the plage region. In Figure~\pref{f5} we show the raw \ion{Ca}{2}H intensity (top left) and background subtracted (bottom left) images at the start of the sequence. The background subtracted images (formed by subtracting a 10 pixel \--$\sim$1.3\arcsec{} square\-- boxcar smoothed image from the original) allow us to isolate the dynamic type-II spicules without time-differencing the data and losing some of their temporal behavior. The background subtraction leaves only structures that are small or thin, such as spicules\footnote{The resulting scale of the background subtracted image is $\pm$50 Data Numbers where the original intensity ranged from 300 to 1500 DN.}, which are easily seen extending outward from the plage region (see online supporting movies). We note the strong visual correspondence between the expansion of the spicules outward, all around the plage region, and the upward extension and morphology of the cool coronal emission and trajectories of the blobs observed in the \trace{} 171\AA{} passband (see Figs.~\pref{f2} and~\pref{f3}). We should note that the type-II spicules in the core of the plage are better isolated using a hi-pass Fourier filtering process \citep{DePontieu2009}, but comparing their apparent velocities with those of the blobs in that location is not straightforward.

The right panel of Fig.~\pref{f5} shows a sample x-t plot from the background-subtracted SOT \ion{Ca}{2}H timeseries (the red dashed lines in the left column panels, offset to approximately the same location as that used in Fig.~\pref{f3} accounting for differential rotation). The panel is speckled with short ($\approx$2\arcsec) bright streaks that are slightly inclined to the horizontal, revealing the typical high velocities of type II spicules, which originate near the small (but strong) magnetic flux concentrations that comprise the plage region. For reference we show a red solid line which indicates a projected velocity of 100km/s. We can easily see that there are many streaks that are faster and not many that are slower, but it appears that there is a range of Type-II spicule velocities consistent with those determined in \citet{DePontieu2007b}. These velocities are, in turn, consistent with those observed with EIS \citep[][]{DePontieu2009}, XRT \citep[][]{Sakao2008} and \trace{} \citep[][]{Schrijver1999}. We also note the rapid recurrence of Type-II all over the plot, but in particular towards the end of the sequence just to the left of the dashed vertical line.

\section{Discussion} 
In the previous section we established the intensity and velocity similarities of blobs observed in image timeseries of the corona from XRT \citep[see also][who call these ``continuous outflows'']{Sakao2008} and TRACE with low-amplitude, high-velocity, upflows revealed by novel spectroscopic diagnosis of emission lines in the upper transition region and corona. We finally connected these blobs to dynamic spicules visible in the upper chromosphere. In each temperature regime we have noted the correspondence in location (footpoints or extending away from them), amplitude (of order a few percent of background intensity) and the range of velocities (of order 100km/s) of these events. Our data is compatible with the following scenario: collections of fine jets initialized in the chromosphere (Type-II spicules) are heated and observed as faint high-velocity upflows at the footpoints of coronal loops (with EIS spectra), which are seen as discrete plasma blobs projecting upward and outward from the footpoints along coronal loops when the angle between the line-of-sight and magnetic field direction increases (XRT/\trace{} images). In other words, we suggest that the flows observed by \citet{Sakao2008} are episodic in nature (blobs), that they are similar to the blobs observed by \trace{} described in \citet{Schrijver1999}, and provide a direct window into the process by which hot plasma is propelled from chromospheric to coronal heights in association with dynamic spicules driven from below \citep{DePontieu2009}.

The ubiquitous presence of faint, but strongly blueshifted plasma at the footpoints of the loop structures that are seen to carry the outflowing blobs provides strong evidence for our interpretation of the data. First, the faint blue component is observed with EIS to be at the level of 1-5\% of the background intensity of the line core. Given the strong upflows of order 100 km/s, and the inevitably changing viewing angle, our scenario thus naturally {\it predicts} that coronal imaging will show faint perturbations that propagate away from the footpoints at velocities of order 100 km/s, with amplitudes of order 1-5\% of the background intensity of the loops \-- exactly what \trace{} and XRT observe. The similarity in velocities, location {\it and} amplitudes provide strong evidence for a connection between all of these phenomena.

The visibility in spectral line profiles of the spicule-associated coronal outflows will strongly depend on the background emission of the loops that carry the flows. For example, along dark, long, coronal loops that have weak background emission, the spicular outflows at high speeds may contribute more to the overall spectral line profile, so that line-centroiding will sample more of the outflowing velocity distribution of the blobs, whereas the line widths will be enhanced. This is exactly what we observe for the blob-carrying loops in the maps of Doppler velocity (line centroid) and line width of the \ion{Fe}{12} 195 \AA\ and \ion{Fe}{14} 264 \AA{} lines (bottom panels of Fig~\pref{f1}). The line centroid outflows of 20-30 km/s in the blob-carrying loops (reported previously by \citet{Harra2007} for the \ion{Fe}{12} line) imply that the physical velocities associated with the RB maps (of order 100 km/s) could be higher by 20-30 km/s. Our analysis suggests that the single Gaussian approach to profile fitting of a line dominated by multiple, unresolved components, will yield an ensemble velocity that, while strong \citep[e.g.,][]{Harra2007, Sakao2008, Doschek2008} does not reflect the full range of velocities revealed by the RB analysis.

The presence of strong spectroscopically determined upflows associated with the blobs sheds light on another issue. \citet{Schrijver1999} interpreted these blobs as episodic flows and suggested that they were driven upward from low heights by Lorentz forces acting in the strong field regions. This scenario fits in well with that proposed by \citet{DePontieu2009}. However, some of these upward propagating disturbances in \trace{} time sequences have been interpreted as propagating slow-mode magneto-acoustic waves \citep[e.g.,][]{deMoortel2002a, deMoortel2002b, McIntosh2008, Wang2009}. Because of the similarities in periods, the waves above plage regions were thought to result from p-mode leakage from the photosphere \citep{DePontieu2005}. The wave interpretation was compatible with the lack of strong Doppler shifts (which would have been expected from significant flows). However, our discovery of high speed (if faint) quasi-periodic upflows at the footpoints suggests that at least a fraction of these ``waves'' may, in fact, be better interpreted as flows. This highlights some of the poorly understood issues regarding p-mode leakage. For example, given the ubiquity of the p-mode leakage process \citep[e.g.,][]{DePontieu2004,Hansteen2006,Jefferies2006,DePontieu2007a,Rouppe2007}, it is surprising that 5 min ``waves'' are seen (with \trace{}) in only a small subset of loops, and usually only on the longest coronal loops \citep[that are faint and cool as a result, e.g.,][]{RTV}. In addition, many of the ``waves'' show, at best, quasi-periodicity and some have periods much longer than the typical p-mode spectrum (10-25 minutes). This begs the question: what portion of the wave-like phenomena observed by \trace{} are the result of p-mode leakage and which are just the result of observations with long exposure times (causing apparent lengthening of the blob) and high local image contrast (because of reduced background emission) of heating events triggered quasi-periodically in the lower atmosphere? The higher signal-to-noise and wide temperature coverage offered by SDO/AIA may help us establish the relative role of waves and episodic flows.

The quasi-periodicity of these events may also provide an intriguing window into the driver of chromospheric spicules (coronal blobs) and ubiquitous Alfv\'{e}nic perturbations of the corona \citep[][]{Tomczyk2007}. Current suggestions for the formation of type-II spicules indicate a role for reconnection \citep{DePontieu2007b} and their rooting in mostly unipolar flux regions may favor some form of ``component'' reconnection mechanism \citep{Langangen2008}. It is tempting to speculate that, whatever the formation mechanism, the quasi-periodicity on timescales of 5-10 minutes may be related to granular dynamics that drive the build-up of magnetic field stresses that lead to the formation of the spicules (and blobs). Work remains to investigate how clumps of concurrent type-II spicules ``merge'' to form the blobs observed by \trace{} and XRT and how the plasma from these apparently discrete heating and injection events merges with the pre-existing coronal plasma.

\section{Conclusion}
We have shown that high velocity upflow events are visible in the chromospheric and coronal footpoints of loops that carry plasma blobs observed by \trace{} and XRT. These velocities correspond spatially and spectrally to weak, blue-wing asymmetries that are observed across a range of temperatures in many EIS lines. We suggest that the weak upflows are related to episodes of dynamic spicule activity. Further, we suggest that the observed coronal outflows seen in long-exposure \trace{} and XRT image sequences of the corona are the result of the heating of chromospheric material to coronal temperatures in dynamic Type-II spicules and, as such, may be the signature of discrete episodic coronal heating and mass injection events. The nature of these Type-II spicules, in tracing out the roots of the coronal magnetic topology and mass loading, suggests a more appropriate (or descriptive) name for these spicules: ``radices'' (singular: ``radix'' \-- Latin for ``root''). 

\acknowledgements
We appreciate discussions with Alan Title and Karel Schrijver. The effort was supported by external funds (SWM: NNX08AL22G, NNX08AU30G  from NASA and ATM-0925177 from the National Science Foundation \-- BDP: NNG06GG79G and NNX08AH45G from NASA), the \hinode{} (NNM07AA01C) and \trace{} NAS5-38099 mission contracts to LMSAL. \hinode{} is a Japanese mission developed and launched by ISAS/JAXA, with NAOJ as a domestic partner and NASA and STFC (UK) as international partners. NCAR is sponsored by the National Science Foundation.

\end{document}